\documentclass[prc,aps,twocolumn]{revtex4}
\usepackage{graphicx}
\usepackage{dcolumn}
\usepackage{bm}

\def\tend{\mathop{\to}}

\def\lim{\mathop{\rm {lim}}}

\begin{document}
\draft \preprint{HEP/123-qed} \widetext
\title{Nonlocality  of the $NN$ interaction in an effective field theory}
\author{Renat Kh.Gainutdinov and Aigul A.Mutygullina}
\address{
Department of Physics, Kazan State University, 18 Kremlevskaya St,
Kazan 420008, Russia }
\email{Renat.Gainutdinov@ksu.ru}
\date{\today}

\begin{abstract}
We investigate low energy nucleon dynamics in the effective field
theory (EFT) of nuclear forces. In  leading order of the
two-nucleon EFT we show that nucleon dynamics is governed by the
generalized dynamical equation with a nonlocal-in-time interaction
operator. This equation is shown to open new possibilities for
applying the EFT approach to the description of low energy nucleon
dynamics.
\end{abstract}
 \maketitle
\narrowtext

\section{Introduction}
\label{sec:level1}

  Understanding how
nuclear forces emerge from the fundamental theory of quantum
chromodynamics (QCD) is one of the most important problems of
quantum physics. To study hadron dynamics at scales where QCD is
strongly coupled, it is useful to employ effective field theories
(EFT's) [1,2], invaluable tools for computing physical
quantities in the theories with disparate energy scales. Following
the early work of Weinberg and others [3-7], the EFT approach has
become very popular in nuclear physics [8,9].  A fundamental
difficulty in an EFT description of nuclear forces is that
effective Lagrangians which are used within its framework yield
graphs which are divergent, and give rise to singular quantum-mechanical
potentials. These potentials do not satisfy the
requirements of ordinary quantum mechanics and need to be
regulated, and renormalization must be performed. In this way one
can successfully perform calculations of many quantities in
nuclear physics. However, in this case one cannot parametrize the
interactions of nucleons, by using some Lagrangian or Hamiltonian.
In fact, in quantum field theory, knowing the Lagrangian is not
sufficient to compute results for physical quantities. In
addition, one needs to specify a way to make all infinite
quantities finite. Another consequence of this situation is that
there are not any equations for renormalized amplitudes in
subtractive EFT's: In order to resolve this problem, in some EFT's
[6,10,11] finite cutoff regularization is used. However,
renormalization is required to render such theories consistent,
and certain cutoff-dependent terms have to be absorbed into the
constants before determining them from empirical data.

If the EFT approach, as it is widely believed, is able to provide
a fundamental description of the interactions of nucleons at low
energies, one can hope that it will give rise to parametrization
of these interactions by interaction operators as
fundamental as the Coulomb potential, that parametrizes the
interaction of charged particles in low energy QED. In the quantum
mechanics of particles interacting via the Coulomb potential, which
is an example of the effective theory, one deals with a well-defined
interaction Hamiltonian and the Schr{\"o}dinger equation governing
the dynamics of the theory. This theory is self-consistent, and
provides an excellent description of atomic phenomena at low
energies. In light of this fact one can expect that the EFT of
nuclear forces will allow one to construct a well-defined operator
parametrizing the interaction of nucleons and governing their
dynamics. At the same time, since the EFT leads to singular
nucleon potentials, this operator must not be an interaction
Hamiltonian. In the present paper we show that recent developments
in quantum theory provide the possibility of a consistent
description of nucleon dynamics predicted by the EFT approach.

The above problem of subtractive EFT's is the same that arises in
any quantum field theory with UV divergences: Regularization and
renormalization allow one to render the physical predictions
finite, however, it is impossible to construct a renormalized
Hamiltonian acting on the Fock space, i.e., after renormalization
the dynamics of the theory is not governed by the Schr{\"o}dinger
equation. This equation is local in time, and the interaction
Hamiltonian describes an instantaneous interaction. On the other
hand, locality is the main cause of UV divergences in quantum
field theory (QFT), and
hence  regularization and renormalization may be considered as
some ways of nonlocalization of the theory. In Ref. [12] it has been
shown that the use of the Feynman approach [13] to quantum theory
in combination with the canonical approach allows one to extend
quantum dynamics to describe the evolution of a system whose
dynamics is generated by a nonlocal-in-time interaction, and an
equation of motion has been derived as the most general dynamical
equation consistent with the current concepts of quantum theory.
Being equivalent to the Schr{\"o}dinger equation in the particular
case where interaction is instantaneous, this equation permits the
generalization to the case where the interaction operator is
nonlocal in time. It has been shown [12] that a generalized
quantum dynamics (GQD) developed in this way provides a new
insight into the problem of  UV divergences.

The aim of the present paper is to show that the formalism of the
GQD developed in Ref. [12] opens new possibilities in the effective
theory of nuclear forces.  We show that in leading order of the
two-nucleon EFT nucleon dynamics is governed by the generalized
dynamical equation with a nonlocal-in-time interaction operator.
By using the example of the $T$ matrix that describes the contact
term in leading order, we investigate the dynamical situation in
an EFT after renormalization. We show that this $T$ matrix has the
properties that are completely unsatisfactory from the point of
view of ordinary quantum mechanics, but satisfies the generalized
dynamical equation. Moreover, in this case we deal with the
dynamics that is described by the model developed in Refs. [12,14]
as a test model demonstrating the possibility of the extension of
quantum dynamics to the case of nonlocal-in-time interactions. As
has been shown [12], there are no physical reasons to restrict
ourselves to the case of local interactions where the generalized
dynamical equation is equivalent to the Schr{\"o}dinger equation:
The dynamics corresponding to any solution of this equation is not
at variance with the current concepts of quantum physics. This
means that the situation where the dynamics is generated by
nonlocal-in-time interaction is possible in principle. In the
present paper we show that this possibility is realized in the
case of low energy nucleon dynamics, and in leading order this
dynamics is governed by a nonlocal-in-time interaction operator
that is considered in the exactly solvable model [12,14]. We will
show that this feature of the GQD permits the parametrization of
the $NN$ interactions by operators that are as well defined as, for
example, the Coulomb potential, and are independent of
renormalization schemes. The advantages of the GQD is that it
allows one to describe the evolution of nucleon systems in a
consistent way by using equations that do not require
renormalization. In the present paper this fact will be proved in
leading order of the EFT approach.

  In Sec. II we
review the principal features of the GQD. In Sec. III we consider
the exactly solvable model with nonlocal-in-time interaction and its
applications to the description  of the  $NN$ interaction. The
dynamics of a quantum system within the EFT is investigated in
Sec. IV.  We will consider the nucleons as spinless particles at
energies much smaller than their mass, their internal excitation
energy, and the range of their interaction, and will investigate
nucleon dynamics within the EFT approach developed in Ref. [15]. We
show that in leading order this dynamics is equivalent to the
dynamics of the model [12,14] with a nonlocal-in-time interaction
operator. The dynamical situation that arises in an EFT after
renormalization is investigated in Sec. V. Finally, in Sec.  VI we
present some concluding  remarks.

\section{Generalized quantum dynamics}
\label{sec:level2} As has been shown in Ref.\cite{R.Kh.:1999}, the
Schr{\"o}dinger equation is not the most general dynamical
equation consistent with the current concepts of quantum theory.
Let us consider these concepts. As is well known, the canonical
formalism is founded on the following assumptions: (i) The
physical state of a system is represented by a vector (properly by
a ray) of a Hilbert space, and (ii) an observable $A$ is
represented by a Hermitian hypermaximal operator $\alpha$. The
eigenvalues $a_r$ of $\alpha$ give the possible values of $A$. An
eigenvector $|\varphi_r^{(s)}\rangle$ corresponding to the
eigenvalue $a_r$ represents a state in which $A$ has the value
$a_r$. If the system is in the state $|\psi\rangle,$ the
probability $P_r$ of finding the value $a_r$ for $A$, when a
measurement is performed, is given by
$$P_{r} = \langle\psi|P_{V_{r}} |\psi\rangle= \sum_s |\langle\varphi_r^{(s)}|\psi\rangle|^2, $$
where $P_{V_{r}}$ is the projection operator on the eigenmanifold
$V_r$ corresponding to $a_r,$ and the sum $\Sigma_s$ is taken over
a complete orthonormal set ${|\varphi_r^{(s)}\rangle}$
($s=1,2,...$) of $V_r.$ The state of the system immediately after
the observation is described by the vector
$P_{V_{r}}|\psi\rangle.$

In canonical formalism these postulates are used together with
the assumption that the time evolution of a state vector is
governed by the Schr{\"o}dinger equation. However, in QFT the
Schr{\"o}dinger equation is only of formal importance because of
the UV divergences. Note in this connection that in the Feynman
approach \cite{Feynman:1948} to quantum theory this equation is
not used as a fundamental dynamical equation. As is well known,
the main postulate, on which this approach is founded, is as
follows:
(iii) The probability of an event is the absolute square of a
complex number called the probability amplitude. The joint
probability amplitude of a time-ordered sequence of events is
product of the separate probability amplitudes of each of these
events. The probability amplitude of an event which can happen in
several different ways is a sum of the probability amplitudes for
each of these ways.

The statements of assumption (iii) express the well-known law
for the quantum-mechanical probabilities. Within canonical
formalism this law is derived as one of the consequences of the
theory. However, in the Feynman formulation of quantum theory this
law is used as the main postulate of the theory. The Feynman
formulation also contains, as its essential idea, the concept of a
probability amplitude associated with a completely specified
motion or path in space-time. From assumption (iii) it then
follows that the probability amplitude of any event is a sum of
the probabilities that a particle has a completely specified path
in space-time. The contribution from a single path is postulated
to be an exponential whose (imaginary) phase is the classical
action (in units of $\hbar$) for the path in question. The above
constitutes the contents of the second postulate of the Feynman
approach to quantum theory. This postulate is not so fundamental
as assumption (iii), which directly follows from the analysis
of the phenomenon of quantum interference \cite{Feynman:1948}. In
Ref.\cite{R.Kh.:1999} it has been shown that the first postulate
of the Feynman approach [assumption (iii)] can be used in
combination with the main fundamental postulates of canonical
formalism [assumptions (i) and (ii)] without resorting to the
second Feynman postulate and the assumption that the dynamics of a
quantum system is governed by the Schr{\"o}dinger equation. Such a
use of the main assumptions of quantum theory leads to a more
general dynamical equation than the Schr{\"o}dinger equation.

In a general case the time evolution of a quantum system is
described by the evolution equation
$$|\Psi(t)\rangle=U(t,t_0)|\Psi(t_0)\rangle,$$ where $U(t,t_0)$ is the unitary
evolution operator,
\begin{equation}
U^{+}(t,t_0) U(t,t_0) = U(t,t_0) U^{+}(t,t_0) = {\bf
1},
\label{unitary}
\end{equation}
with the group property
\begin{equation}
U(t,t') U(t',t_0) = U(t,t_0), \quad U(t_0,t_0) ={\bf 1}.
\label{compos}
\end{equation}
Here we use the interaction picture. According to assumption
(iii), the probability amplitude of an event which can happen in
several different ways is a sum of contributions from each
alternative way. In particular, the amplitude
 $\langle\psi_2| U(t,t_0)|\psi_1\rangle$ can be represented as a sum
of contributions from all alternative ways of realization of the
corresponding evolution process. Dividing these alternatives into
different classes, we can then analyze such a probability
amplitude in different ways. For example, subprocesses with
definite instants of the beginning and  end of the interaction in
the system can be considered as such alternatives. In this way the
amplitude $\langle\psi_2|U(t,t_0)|\psi_1\rangle$  can be written
in the form \cite{R.Kh.:1999}
\begin{eqnarray}
\langle\psi_2| U(t,t_0)|\psi_1\rangle = \langle\psi_2|\psi_1\rangle \nonumber\\
+\int_{t_0}^t dt_2 \int_{t_0}^{t_2} dt_1 \langle\psi_2|\tilde
S(t_2,t_1)|\psi_1\rangle, \label{repre}
\end{eqnarray}
where $\langle\psi_2|\tilde S(t_2,t_1)|\psi_1\rangle$ is the
probability amplitude that if at time $t_1$ the system was in the
state $|\psi_1\rangle,$ then the interaction in the system will
begin at time $t_1$ and will end at  time $t_2,$ and at this time
the system will be in the state $|\psi_2\rangle$. Note that in
general $\tilde S(t_2,t_1)$  may be only an operator-valued
generalized function of $t_1$ and $t_2$, since only $U(t,t_0)={\bf
1}+ \int^{t}_{t_0} dt_2 \int^{t_2}_{t_0}dt_1\tilde S(t_2,t_1)$
must be an operator on the Hilbert space. Nevertheless, it is
convenient to call $\tilde S(t_2,t_1)$ an "operator", using this
word in the generalized sense. In the case of an isolated system
the operator $\tilde S(t_2,t_1)$ can be represented in the form
\cite{R.Kh.:1999}
\begin{equation}
\tilde S(t_2,t_1) = \exp(iH_0t_2) \tilde T(t_2-t_1) \exp(-iH_0 t_1),
\label{t}
\end{equation}
with $H_0$ being the free Hamiltonian.

As has been shown in Ref.\cite{R.Kh.:1999}, for the evolution
operator $U(t,t_0)$ given by Eq. (\ref{repre}) to be unitary for any
times $t_0$ and $t$, the operator $\tilde S(t_2,t_1)$ must satisfy
the following equation:
\begin{eqnarray}
(t_2-t_1) \tilde S(t_2,t_1) = \int^{t_2}_{t_1} dt_4
\int^{t_4}_{t_1}dt_3 \nonumber \\
 \times(t_4-t_3) \tilde S(t_2,t_4) \tilde S(t_3,t_1).
\label{main}
\end{eqnarray}
This equation allows one to obtain the operators $\tilde
S(t_2,t_1)$ for any $t_1$ and $t_2$, if the operators $\tilde
S(t'_2, t'_1)$ corresponding to infinitesimal duration times $\tau
= t'_2 -t'_1$ of interaction are known. It is natural to assume
that most of the contribution to the evolution operator in the
limit $t_2 \to t_1$ comes from the processes associated with the
fundamental interaction in the system under study. Denoting this
contribution by $H_{int}(t_2,t_1)$, we can write
\begin{equation}
\tilde{S}(t_2,t_1) \tend\limits_{t_2\rightarrow t_1}
H_{int}(t_2,t_1) + o(\tau^{\epsilon}), \label{bound}
\end{equation}
where $\tau=t_2-t_1$. The parameter $\varepsilon$ is determined by
demanding that $H_{int}(t_2,t_1)$ must be so close to the solution
of Eq. (\ref{main}) in the limit $t_2\tend t_1$ that this equation
has a unique solution having the behavior (\ref{bound}) near the
point $t_2=t_1$. Thus this operator must satisfy the condition
\begin{eqnarray}
(t_2-t_1) H_{int}(t_2,t_1)\tend\limits_{t_2 \tend t_1}
\int^{t_2}_{t_1} dt_4 \int^{t_4}_{t_1} dt_3 (t_4-t_3)\nonumber\\
 \times
H_{int}(t_2,t_4) H_{int}(t_3,t_1)+ o(\tau^{\epsilon+1}).
\label{bound'}
\end{eqnarray}
Note that the value of the parameter $\epsilon$ depends on the
form of the operator $ H_{int}(t_2,t_1).$ Since $\tilde
S(t_2,t_1)$ and $H_{int}(t_2,t_1)$ are only operator-valued
distributions, the mathematical meaning of the conditions
(\ref{bound}) and (\ref{bound'}) needs to be clarified. We will
assume that condition (\ref{bound}) means that
\begin{eqnarray}
\langle\Psi_2|\int^{t}_{t_0} dt_2 \int^{t_2}_{t_0}dt_1\tilde
S(t_2,t_1)
|\Psi_1\rangle\nonumber\\
\tend\limits_{t\tend t_0} \langle\Psi_2|\int^{t}_{t_0} dt_2
\int^{t_2}_{t_0}dt_1
H_{int}(t_2,t_1)|\Psi_1\rangle+o(\tau^{\epsilon+2}), \nonumber
\end{eqnarray}
 for
any vectors $|\Psi_1\rangle$ and $|\Psi_2\rangle$ of the Hilbert
space. Condition (\ref{bound'}) has to be  considered in the same
sense.

Within GQD the operator $H_{int}(t_2,t_1)$ plays the role
the interaction Hamiltonian plays in the ordinary
formulation of quantum theory: It generates the dynamics of a
system. Being a generalization of the interaction Hamiltonian,
this operator is called the generalized interaction operator. If
$H_{int}(t_2,t_1)$ is specified, Eq. (\ref{main}) allows one to
find the operator $\tilde S(t_2,t_1).$ Formula (\ref{repre}) can
then be used to construct the evolution operator $U(t,t_0)$ and
accordingly the state vector,
\begin{eqnarray}
|\psi(t)\rangle = |\psi(t_0)\rangle +  \int_{t_0}^t dt_2
\int_{t_0}^{t_2} dt_1 \tilde S(t_2,t_1) |\psi(t_0)\rangle,
\label{psi}
\end{eqnarray}
 at any time $t.$ Thus
Eq. (\ref{main}) can be regarded as an equation of motion for
states of a quantum system. By using Eqs. (\ref{repre}) and (\ref{t}),
the evolution operator can be represented in the form
\begin{eqnarray}
\langle n_2|U(t,t_0)|n_1\rangle= \langle
n_2|n_1\rangle+\frac{i}{2\pi} \int^\infty_{-\infty}
dx\label{evolution}\nonumber\\
 \times \frac
{\exp[-i(z-E_{n_2})t] \exp[i(z-E_{n_1})t_0]}
{(z-E_{n_2})(z-E_{n_1})} \nonumber\\
 \times\langle n_2|T(z)|n_1\rangle,
\end{eqnarray}
 where $z=x+iy$, $y\rangle0$, and
\begin{equation}
\langle n_2|T(z)|n_1\rangle = i \int_{0}^{\infty} d\tau
\exp(iz\tau) \langle n_2|\tilde T(\tau)|n_1\rangle. \label{tt}
\end{equation}
From Eq. (\ref{evolution}), for the evolution operator in the
Schr{\"o}dinger picture, we get
\begin{equation}
U_s(t,0)=\frac{i}{2\pi}\int^\infty_{-\infty} dx\exp
(-izt)G(z),\label{schr}
\end{equation}
where
\begin{equation}
\langle n_2|G(z)|n_1\rangle=\frac{\langle
n_2|n_1\rangle}{z-E_{n_1}}+ \frac{\langle
n_2|T(z)|n_1\rangle}{(z-E_{n_2})(z-E_{n_1})}.\label{lventa}
\end{equation}
Eq. (11) is the well-known expression establishing the connection
between the evolution operator and the Green operator $G(z)$, and
can be regarded as the definition of the operator $G(z)$.

The equation of motion (\ref{main}) is equivalent to the following
equation for the $T$ matrix \cite{R.Kh.:1999}:
\begin{equation}
\frac{d\langle n_2|T(z)|n_1\rangle }{dz} \nonumber\\
=- \sum \limits_{n}\frac{\langle n_2|T(z)|n\rangle\langle
n|T(z)|n_1\rangle}{(z-E_n)^2}, \label{dif}
\end{equation}
with the boundary condition
\begin{equation}
\langle n_2|T(z)|n_1\rangle \tend \limits_{|z| \tend \infty}
\langle n_2| B(z)|n_1\rangle+o(|z|^{-\beta}),\label{dbound}
\end{equation}
where
$$
B(z) = i \int_0^{\infty} d\tau \exp(iz \tau) H^{(s)}_{int}(\tau),
$$
$\beta=1+\epsilon,$ and $$H^{(s)}_{int}(t_2-t_1) =
 \exp(-iH_0t_2)H_{int}(t_2,t_1) \exp(iH_0t_1)$$
 is the generalized interaction
operator in the Schr{\"o}dinger picture. The solution of
Eq. (\ref{dif}) satisfies the equation
\begin{eqnarray}
\langle n_2|T(z_1)|n_1\rangle - \langle n_2|T(z_2)|n_1\rangle  \nonumber \\
=(z_2 -z_1)
\sum_n
 \frac {\langle n_2|T(z_2)|n\rangle\langle n|T(z_1)|n_1\rangle}
{(z_2-E_n)(z_1-E_n)}. \label{difer}
\end{eqnarray}
This equation in turn is equivalent to the following equation for
the Green operator:
\begin{equation}
G(z_1)-G(z_2)=(z_2-z_1)G(z_2)G(z_1).\label{green}
\end{equation}
This is the  Hilbert identity, which in the Hamiltonian formalism
follows from the fact that in this case the evolution operator
(11) satisfies the Schr{\"o}dinger equation, and hence the Green
operator is of the form
\begin{equation}
G(z)=(z-H)^{-1},\label{reso}
\end{equation}
with $H$ being the total Hamiltonian.  At the same time, as has been
shown in Ref.\cite{R.Kh.:1999}, Eq. (\ref{main}) and hence
Eqs. (\ref{dif}) and (\ref{difer}) are unique consequences of the
unitarity condition and the representation Eq. (\ref{repre}),
expressing the Feynman superposition principle [assumption
(iii)]. It should be noted that the evolution operator constructed
by using the Schr{\"o}dinger equation can be represented [16] in
the form (\ref{repre}). Being written in terms of the operators
$\tilde S(t_2,t_1)$, Eq. (\ref{main}) does not contain operators
describing interaction in the system. It is a relation for $\tilde
S(t_2,t_1)$, which are the contributions to the evolution operator
from the processes with defined instants of the beginning and end
of the interaction in the systems. Correspondingly,
Eqs. (\ref{dif}) and (\ref{difer}) are relations for the $T$ matrix.
A remarkable feature of  Eq. (\ref{main}) is that it works as a
recurrence relation, and to construct the evolution operator it is
sufficient to know the contributions to this operator from the
processes with infinitesimal duration times of interaction, i.e.
from the processes of a fundamental interaction in the system.
This makes it possible to use Eq. (\ref{main}) as a dynamical
equation. Its form does not depend on the specific feature of the
interaction (the Schr{\"o}dinger equation, for example, contains
the interaction Hamiltonian). Since Eq. (\ref{main}) must be
satisfied in all the cases, it can be considered as the most
general dynamical equation consistent with the current concepts of
quantum theory. All the needed dynamical information is contained in
the boundary condition for this equation, i.e., in the generalized
interaction operator $H_{int}(t_2,t_1)$. As has been shown in
Ref. \cite{R.Kh.:1999}, the dynamics governed by Eq. (\ref{main}) is
equivalent to the Hamiltonian dynamics in the case where the
operator $H_{int}(t_2,t_1)$ is of the form
\begin{equation}
 H_{int}(t_2,t_1) = - 2i \delta(t_2-t_1)
 H_{I}(t_1) ,
 \label{delta}
\end{equation}
with $H_{I}(t_1)$ being the interaction Hamiltonian in the
interaction picture. In this case the state vector
$|\psi(t)\rangle$ given by Eq. (\ref{psi}) satisfies the
Schr{\"o}dinger equation,
$$
  \frac {d |\psi(t)\rangle}{d t} = -iH_I(t)|\psi(t)\rangle.
$$
The delta function $\delta(\tau)$ in Eq. (\ref{delta}) emphasizes that
in this case the fundamental interaction is instantaneous. Thus
the Schr{\"o}dinger equation results from the generalized equation
of motion (\ref{main}) in the case where the interaction
generating the dynamics of a quantum system is instantaneous. At
the same time, Eq. (\ref{main}) permits the generalization to the
case where the interaction generating the dynamics of a quantum
system is nonlocal in time \cite{R.Kh.:1999,R.Kh./A.A.:1999}. In
a general case, the generalized interaction operator has the
following form \cite{R.Kh./A.A.:1999}:
$$H_{int}(t_2,t_1)=-2i\delta(t_2-t_1)H_I(t_1)+H_{non}(t_2,t_1),$$
where the first term on the right-hand side of this equation
describes the instantaneous component of the interaction
generating the dynamics of a quantum system, while the term
$H_{non}(t_2,t_1)$ represents its nonlocal-in-time part. As has
been shown, there is one-to-one correspondence between nonlocality
of interaction and the UV behavior of the matrix elements of the
evolution operator as a function of momenta of particles: The
interaction operator can be nonlocal in time only in the case
where this behavior is "bad", i.e., in a local theory it results in
UV divergences. In Ref. \cite{R.Kh.:2001} it has been shown that
after renormalization the dynamics of the three-dimensional theory
of a neutral scalar field interacting through a $\varphi^4$
coupling is governed by the generalized dynamical Eq.
(\ref{main}) with a nonlocal-in-time interaction operator. This
lets us expect that after renormalization the dynamics of an EFT is
also governed by this equation with a nonlocal interaction
operator. In the next section we will consider this problem by
using a toy model of the separable $NN$ interaction.

\section{Models with nonlocal-in-time interactions
and the short-range  $NN$ interaction} Let us consider the evolution
problem for two nonrelativistic particles in the center-of-mass. We denote
the relative momentum by ${\bf p}$ and the nucleon mass by $m$.
Assume that the generalized interaction operator in the
Schr{\"o}dinger picture $H^{(s)}_{int}(\tau)$ has the form
$$
\langle{\bf p}_2| H^{(s)}_{int}(\tau) |{\bf p}_1\rangle =
\varphi^*({\bf p}_2) \varphi({\bf p}_1) f(\tau),
$$
where $f(\tau)$ is some function of $\tau,$ and the form factor
$\varphi ({\bf p})$ has the following asymptotic behavior for
$|{\bf p}| \tend \infty:$
\begin{equation}
\varphi({\bf p}) \sim |{\bf p}|^{-\alpha}, \quad {(|{\bf p}| \tend
\infty).} \label{form}
\end{equation}
Let, for example, $\varphi ({\bf p})$ be of the form
$$
\varphi({\bf p}) = |{\bf p}|^{-\alpha}+g({\bf {p}}),
$$
and in the limit $|{\bf p}| \tend \infty$ the function $g({\bf
{p}})$ satisfies the estimate $g({\bf {p}})=o(|{\bf
{p}}|^{-\delta})$, where $\delta>\alpha,$ $\delta>\frac{3}{2}.$ In
the separable case, $\langle{\bf p}_2| \tilde S(t_2,t_1) |{\bf
p}_1\rangle$ can be represented in the form
$$
\langle{\bf p}_2| \tilde S(t_2,t_1) |{\bf
p}_1\rangle=\varphi^*({\bf p}_2)\varphi({\bf p}_1)\tilde
s(t_2,t_1).
$$
Correspondingly,
 $\langle{\bf p}_2| T(z) |{\bf p}_1\rangle$ is of the form
\begin{equation}
  \langle{\bf p}_2| T(z)|{\bf p}_1\rangle = \varphi^* ({\bf p}_2)\varphi ({\bf p}_1)
t(z),\label{separ}
\end{equation}
 where, as it follows from Eq. (\ref{dif}), the function $t(z)$
satisfies the equation
\begin{equation}
\frac {dt(z)}{dz} = -t^2(z) \int \frac{d^3k}{(2\pi)^3} \frac
{|\varphi ({\bf k})|^2} {(z-E_k)^2} \label{deq}
\end{equation}
with the asymptotic condition
\begin{equation}
t(z)  \tend \limits_{|z| \tend \infty} f_1(z) + o(|z|^{-\beta}),
\label{asym}
\end{equation}
where
\begin{equation}
f_1(z)= i \int_0^{\infty} d\tau \exp(iz\tau) f(\tau), \label{fn}
\end{equation}
 and $E_k =
\frac {k^2}{2 \mu}$. The solution of Eq. (\ref{deq}) with the
initial condition $t(a)=g_a,$ where $a \in (-\infty,0),$ is
\begin{equation}
t(z)\nonumber\\
 = g_a \left(1 +(z-a) g_a
 \int \frac{d^3k}{(2\pi)^3} \frac {|\varphi ({\bf k})|^2}
{(z-E_k)(a-E_k)} \right)^{-1}.\label{deqa}
\end{equation}
In the case $\alpha >\frac{1}{2}$, the function $t(z)$ tends to a
constant as $|z| \tend \infty$:
\begin{equation}
t(z)  \tend \limits_{|z| \tend \infty} \lambda.\label{lambda}
\end{equation}
Thus in this case the function $f_1(z)$ must also tend to
$\lambda$ as $|z| \tend \infty.$ From this it follows that the
only possible form of the function $f(\tau)$ is
$$
f(\tau) = -2i \lambda \delta(\tau) + f^{\prime}(\tau),
$$
where the function $f^{\prime}(\tau)$ has no such a singularity at
the point $\tau=0$ as does the delta function. In this case  the
generalized interaction operator $H^{(s)}_{int}(\tau)$ has the
form
\begin{equation}
\langle{\bf p}_2| H^{(s)}_{int}(\tau) |{\bf p}_1\rangle=-2i
\lambda \delta(\tau)\varphi^*({\bf p}_2) \varphi({\bf
p}_1),\label{instant}
\end{equation}
 and hence the dynamics generated by this operator is
equivalent to the dynamics governed by the Schr{\"o}dinger
equation with the separable potential
\begin{equation}
\langle{\bf p}_2|H_I|{\bf p}_1\rangle = \lambda \varphi^*({\bf
p}_2) \varphi({\bf p}_1).\label{ham}
\end{equation}
Solving Eq. (\ref{deq}) with the boundary condition (\ref{lambda}),
we easily get the well-known expression for the $T$ matrix in the
separable-potential model
\begin{equation}
\langle{\bf p}_2|T(z)|{\bf p}_1\rangle \nonumber\\
= \varphi^* ({\bf p}_2)\varphi({\bf
p}_1) \left (\frac{1}{\lambda} - \int \frac{d^3k}{(2\pi)^3} \frac
{|\varphi({\bf k})|^2}{z-E_k} \right )^{-1}.\label{sol}
\end{equation}

Ordinary quantum mechanics does not permit the extension of the
above model to the case $\alpha \leq \frac{1}{2}.$ Indeed, in the
case of such a large-momentum behavior of the form factors
$\varphi({\bf p}),$ the use of the interaction Hamiltonian given
by Eq. (\ref{ham}) leads to the UV divergences, i.e., the integral
in (\ref{sol}) is not convergent. We will now show that the
generalized dynamical Eq. (\ref{main}) allows one to extend this
model to the case $-\frac{1}{2} < \alpha\leq\frac{1}{2}.$ Let us
determine the class of the functions $f_1(z)$ and correspondingly
the value of $\beta$ for which Eq. (\ref{deq}) has a unique
solution having the asymptotic behavior (\ref{asym}). In the case
$\alpha <\frac{1}{2},$ the function $t(z)$ given by (\ref{deqa})
has the following behavior for $|z| \tend \infty:$
\begin{equation}
t(z)  \tend \limits_{|z| \tend \infty}  b_1
(-z)^{\alpha-\frac{1}{2}}+ b_2 (-z)^{2 \alpha-1} + o(|z|^{2
\alpha-1}),\label{tzero}
\end{equation}
where $b_1 =- 4\pi \cos(\alpha \pi) m^{\alpha-\frac{3}{2}}$ and
$b_2= b_1 |a|^{\frac{1}{2}- \alpha} -b_1^2[M(a)+g_a^{-1}]$ with
$$
M(a) = \int\frac{d^3k}{(2\pi)^3} \frac {|\varphi({\bf k})|^2-
 |{\bf {k}}|^{-2\alpha}}
{a-E_k} .
$$
The parameter $b_1$ does not depend on $g_a.$ This means that all
solutions of Eq. (\ref{deq}) have the same leading term  in Eq. (29),
and only the second term distinguishes the different solutions of
this equation. Thus, in order to obtain a unique solution of
Eq. (\ref{deq}), we must specify the first two terms in the
asymptotic behavior of $t(z)$ for $|z| \tend  \infty.$ From this
it follows that the functions $f_1(z)$ must be of the form
$$
f_1(z) = b_1 (-z)^{\alpha-\frac{1}{2}} + b_2 (-z)^{2 \alpha -1} ,
$$
and $\beta=1-2 \alpha.$ Correspondingly, the functions $f(\tau)$
must be of the form
\begin{equation}
f(\tau) = a_1 \tau^{-\alpha-\frac{1}{2}} + a_2 \tau^{-2 \alpha},
\label{ft}
\end{equation}
with $a_1= -ib_1 \Gamma ^{-1}(\frac{1}{2}-\alpha)
\exp[i(-\frac{\alpha}{2}+ \frac{1}{4}) \pi],$ and $a_2= b_2 \Gamma
^{-1}(1-2\alpha) \exp(-i \alpha \pi),$ where $\Gamma(z)$ is the
gamma function. This means that in the case $\alpha <\frac{1}{2}$
the generalized interaction operator must be of the form
\begin{eqnarray}
\langle{\bf p}_2| {H}^{(s)}_{int}(\tau)|{\bf p}_1\rangle
 = \varphi^* ({\bf p}_2)\varphi ({\bf p}_1)\nonumber\\
\times \left( a_1\tau^{-\alpha-\frac{1}{2}} + a_2
\tau^{-2 \alpha}\right), \label{less}
\end{eqnarray}
and, as it follows from Eqs. (\ref{separ}) and (24), for the $T$ matrix
we have
\begin{equation}
\langle{\bf p}_2| T(z)|{\bf p}_1\rangle = N(z) \varphi^* ({\bf
p}_2)\varphi ({\bf p}_1),\label{nz}
\end{equation}
with
$$
N(z) = g_a \left (1 + (z-a) g_a \int \frac{d^3 k}{(2\pi)^3}
\frac{|\varphi({\bf k})|^2} {(z-E_k)(a- E_k)} \right )^{-1},
$$
where $$ g_a = b_1^2\left (b_1 |a|^{\frac{1}{2} -\alpha} -b_2
-b_1^2M(a)\right )^{-1}. $$ It can be easily checked that $N(z)$
can be represented in the following form
$$
N(z)=\frac{b_1^2}{-b_2+b_1(-z)^{\frac{1}{2}-\alpha}+M(z)b_1^2}.
$$

By using Eqs. (\ref{evolution}) and (\ref{nz}), we can construct the
evolution operator
\begin{eqnarray}
\langle{\bf p}_2|U(t,t_0)|{\bf p}_1\rangle = \langle{\bf p}_2|{\bf
p}_1\rangle + \frac
{i}{2\pi} \int_{-\infty}^{\infty} dx \nonumber\\
\times \frac {\exp[-i(z-E_{p_2})t]
\exp[i (z - E_{p_1})t_0]} {(z-E_{p_2})(z-E_{p_1})} \nonumber\\
\times \varphi^*({\bf p}_2) \varphi({\bf p}_1)N(z), \label{evol}
\end{eqnarray}
 where
$z=x+iy,$ and $y>0.$ The evolution operator $U(t,t_0)$ defined by
Eq. (\ref{evol}) is a unitary operator satisfying the composition
law (\ref{compos}), provided that the parameter $b_2$ is real.

In the case $\alpha =\frac{1}{2}$
 the generalized interaction operator must be of the form [17]
\begin{eqnarray}
 \langle{\bf p}_2| {H}^{(s)}_{int}(\tau)|{\bf p}_1\rangle = -\frac{i}{2\pi}
\varphi^* ({\bf p}_2) \varphi ({\bf
 p}_1)\nonumber \\
 \times\int_{-\infty}^{\infty}dx
 \exp(-iz\tau)\left(\frac{b_1}{\ln(-z)}+\frac{b_2}{
 \ln^{2}(-z)}\right),\label{equal}
\end{eqnarray}
 where $b_1 =- \frac{4\pi^2}{m}$.

In Ref. \cite{R.Kh./A.A.:1999} the model for $\alpha<\frac{1}{2}$
was used for describing the $NN$ interaction at low energies. The
motivation to use such a model for parametrization of the $NN$
forces is the fact that due to the quark and gluon degrees of
freedom the $NN$ interaction must be nonlocal in time. However,
because of the separation of scales the system of hadrons should
be regarded as a closed system, i.e., the evolution operator must
be unitary and satisfy the composition law. From this it follows
that the dynamics of such a system is governed by Eq. (\ref{main})
with a nonlocal-in-time interaction operator. Let us now
parametrize the $NN$ interaction by the generalized interaction
operator of the form Eq. (31) with the following form factor:
$$
\varphi({\bf p})=\chi({\bf p})+c_1 g_Y({\bf p}), $$ with
$$\chi({\bf p})=\left(d^2+ p^2\right)^{-\frac{\alpha}{2}},
\quad -\frac{1}{2}<\alpha<\frac{1}{2},$$ with $g_Y({\bf p})$ being
the Yamaguchi form factor,
$$
g_Y({\bf p})=\frac{1}{\beta^2+p^2}. $$ Here $d,$ $c_1$, and
$\beta$ are some constants. As it is the generalization of the
Yamaguchi model \cite{Yam} to the case where the $NN$ interaction
is nonlocal in time, our model yields the $NN$ phase shifts in
good agreement with experiment (see Figs.{1-3}). The parameters of
the model are quoted in Table I. However, the main advantage of
this model is that it allows one to investigate the effects of the
retardation in the $NN$ interaction caused by the existence of the
quark and gluon degrees of freedom on nucleon dynamics.

\begin{table}
\caption{The parameters of the interaction operator obtained  by
fitting the $NN$ data,  $\rho=1 {\text MeV}^{-1}$.}
\begin{tabular}{|c|c|c|c|c|c|}
\tableline partial wave&  $\alpha$ &$c_1$ & $\beta\times \rho$ &
$d\times \rho$ & $ b_2\times \rho^{1-2\alpha}$ \\
\tableline ${}^3S_1(np)$ & 0.499 & $133.5\times 10^2$ &433.8 &
$766.2$
& $4.207\times 10^{-5}$  \\
\tableline ${}^1S_0(np)$ & 0.499 & 131.8 & 356.3 & $3.651\times
10^6$  & $4.202\times 10^{-5}$ \\
\tableline ${}^1S_0(pp)$ & 0.499 & 320.0 & 371.7 & $6.763\times
10^5$  & $4.204\times 10^{-5}$ \\
\tableline
\end{tabular}
\end{table}

\begin{figure}
\resizebox{1\columnwidth}{!}{\includegraphics{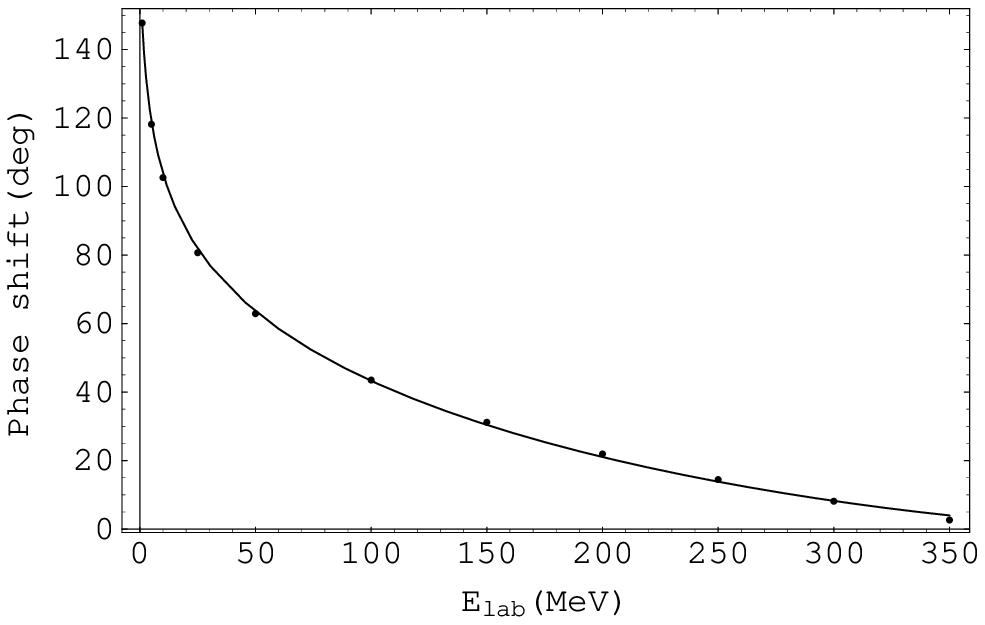}}
\caption{Phase shifts (solid line) in the ${}^3S_1$ channel for $np$
scattering, compared to the experimental data (points)
(Ref. \cite{Stoks:1993}).}
\end{figure}
\begin{figure}
\resizebox{1\columnwidth}{!}{\includegraphics{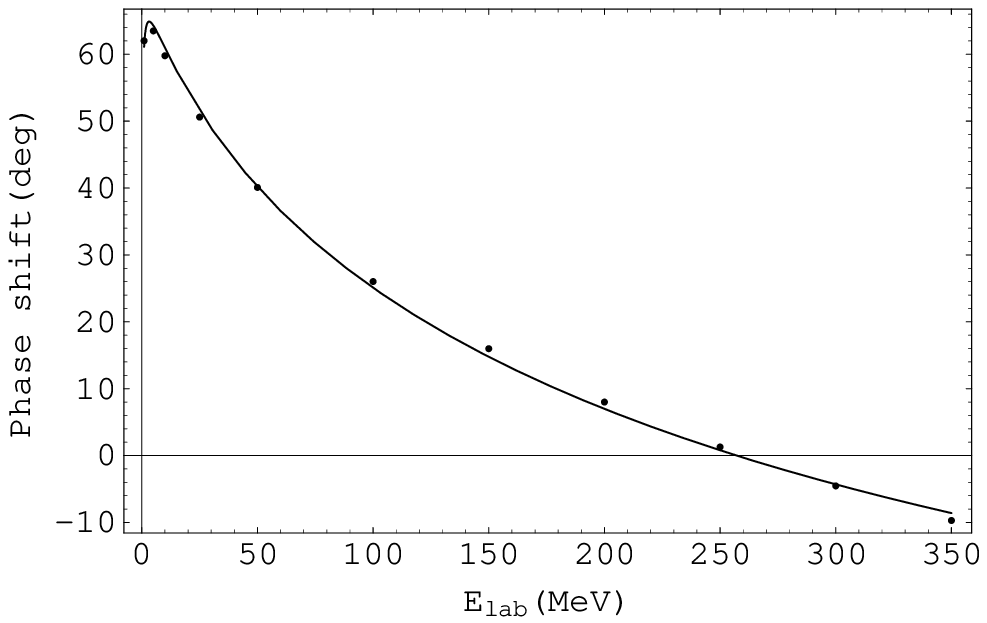}}
\caption{Phase shifts (solid line) in the ${}^1S_0$ channel for
$np$ scattering, compared to the experimental data (points) (Ref.
\cite{Stoks:1993}).}
\end{figure}
\begin{figure}
\resizebox{1\columnwidth}{!}{\includegraphics{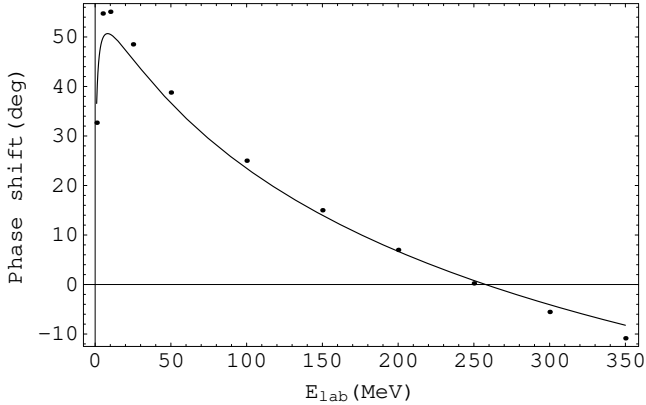}}
\caption{Phase shifts (solid line) in the ${}^1S_0$ channel for $pp$
scattering, compared to the experimental data (points)
(Ref. \cite{Stoks:1993}).}
\end{figure}

As we have seen, there is the one-to-one correspondence between
the form of the generalized interaction operator and the UV
behavior of the form factor $\varphi({\bf p}).$ In the case
$\alpha >\frac{1}{2},$ the operator $H^{(s)}_{int}(\tau)$ must
necessarily have the form (\ref{instant}). In this case the
fundamental interaction is instantaneous. In the case
$-\frac{1}{2} <\alpha <\frac{1}{2}$ [the restriction $\alpha
>-\frac{1}{2}$ is necessary for the integral in Eq. (\ref{deqa}) to
be convergent], the only possible form of $H^{(s)}_{int}(\tau)$ is
Eq. (\ref{less}), and, in the case $\alpha=\frac{1}{2}$, it must
be of the form (\ref{equal}), and hence the interaction generating
the dynamics of the system is nonlocal in time. Thus the
interaction generating the dynamics can be nonlocal in time only
if the form factors have the "bad" large-momentum behavior that
within Hamiltonian dynamics gives rise to the ultraviolet
divergences:
\begin{mathletters}
\begin{eqnarray}
\text{locality}\quad\Leftrightarrow\varphi({\bf p})\quad\sim
\quad|{\bf
p}|^{-\alpha},\quad \alpha>\frac{1}{2}, \nonumber\\
\text{nonlocality}\Leftrightarrow\varphi({\bf
p})\quad\sim\quad|{\bf p}|^{-\alpha},\quad
\alpha\leq\frac{1}{2}.\nonumber
\end{eqnarray}
\end{mathletters}
From this it follows that the quark-gluon retardation effects must
result in the "bad" UV behavior of the matrix elements of the
evolution operator as a function of momenta of hadrons. Note that
EFT's lead to the same conclusion: Within the EFT approach the
quark and gluon degrees of freedom manifest themselves in the form
of Lagrangians consistent with the symmetries of QCD which gives
rise to the UV divergences. Note also that EFT's are local
theories, despite the existence of the external quark and gluon
degrees of freedom. However, renormalization of EFT's gives rise
to the fact that these theories become nonlocal.

\section{The EFT approach and the nonlocality in time  of
effective interactions}

Let us now investigate the character of the dynamics generated by
the two-nucleon EFT. For the sake of simplicity we will restrict
ourselves to the model in which the nucleons are considered as
spinless. Within the EFT approach the nucleons are considered as
particles (described by a field $\Psi$) with three-momenta $Q$ much
smaller than their mass $m$, the mass difference $\Delta$ to their
first excited state, and the pion mass. If momenta are also small
compared to the range of interaction, the $NN$ interaction can be
approximated by a sequence of contact interactions, with an
increasing number of derivatives. Restricting to parity and
time-reversal theories the effective Lagrangian can be written as
[15]
\begin{eqnarray}
{\cal L}=\Psi^+\left(i\partial_0+\frac{1}{2m}\nabla^2+\frac{1}{8m^3}
\nabla^4+\ldots\right)\Psi
\nonumber\\-\frac{1}{2}C_0\Psi^+\Psi\Psi^+\Psi
-\frac{1}{8}(C_2+C_2')
\{\Psi^+(\overrightarrow{\nabla}\nonumber\\-
\overleftarrow{\nabla})\Psi \cdot\Psi^+\left(\overrightarrow{\nabla}-
\overleftarrow{\nabla}\right)\Psi
-\Psi^+\Psi\Psi^+\left(\overrightarrow{\nabla}-
\overleftarrow{\nabla}\right)^2\Psi\}\nonumber\\
+\frac{1}{8}(C_2-C_2')\Psi^+\Psi
\nabla^2(\Psi^+\Psi)+\ldots,\label{lagrang}
\end{eqnarray}
where $C_{2n}'$s are parameters that depend on the details of the
dynamics of range $\sim\frac{1}{m}$. The particles $\Psi$ are
nonrelativistic, and evolve only forward in time. Particles
number is also conserved. Canonical quantization leads to familiar
Feynman rules, and the $\Psi$ propagator at four-momenta $p$ is
given by
\begin{equation}
S(p^0,{\bf p})=\frac{i}{p^0-\frac{{\bf p}^2}{2m}+\frac{{\bf
p}^4}{8m^3}+\ldots+i\varepsilon}.\label{prop}
\end{equation}
The four-$\Psi$ contact interaction is given by $-iv(p,p')$, with
\begin{equation}
v(p,p')=C_0+C_2({\bf p}^2+{\bf p'}^2)+2C_2'{\bf p}\cdot{\bf
p'}+\ldots,\label{pot}
\end{equation}
with ${\bf p} ({\bf p'})$ the relative momentum of the incoming
(outgoing) particles. The problem can also be easily solved by
using the time-ordered perturbation theory, since in this case we
deal only with the particles evolving forward in time. Obviously
this way is more convenient for constructing the off-shell T
matrix.

Let us consider the two-particle system at energy $E=\frac{{\bf
k}^2}{m}-\frac{{\bf k}^4}{4m^3}+\ldots$ in the center-of-mass
frame. The key point of the EFT approach is that the problem can
be solved by expanding in the number of derivatives at the
vertices or particle lines. In leading order one has to keep only
the first term $C_0$ in Eq. (\ref{pot}), and correspondingly only the
first two terms $p^0$ and $\frac{{\bf p}^2}{2m}$ in the propagator
(\ref{prop}). In this case the two particles evolve according to
the familiar nonrelativistic Schr{\"o}dinger propagator
\begin{equation}
G_0(z)=\int \frac{d^3k}{(2\pi)^3}\frac{|{\bf k}\rangle\langle{\bf
k}|}{z-\frac{{\bf k}^2}{m}+i\varepsilon}.
\end{equation}
In this order the Lagrangian can be rewritten in the form
\begin{eqnarray}
{\cal L}=\Psi^+\left(i\partial_0+\frac{1}{2m}\nabla^2\right)\Psi
-\frac{1}{2}C_0\Psi^+\Psi\Psi^+\Psi.\label{lagr}
\end{eqnarray}
Conservation of particle number reduces the two-nucleon $T$ matrix
to a sum of bubble diagrams. The ultraviolet divergencies can all
be absorbed in the renormalized parameter $C_0^{(R)}$. Summing the
bubbles to a geometric series, one gets the $T$ matrix \cite{Kolck}
\begin{equation}
\langle{\bf p}_2|T^{(0)}(z)|{\bf
p}_1\rangle=-\left[\frac{1}{C_0^{(R)}}+\frac{im^{
\frac{3}{2}}{\sqrt{z}}}{4\pi}\right]^{-1}.
\end{equation}
Note that one can obtain the same $T$ matrix via the $LS$ equation
with the potential $V_0({\bf p}_2,{\bf p}_1)=\widetilde{C}$, by
using some regularization and renormalization procedures.

Let us now consider this problem from the point of view of the
GQD. As we have stated, the character of the dynamics of the
system depends on the large-momentum behavior of the matrix
elements of the evolution operator. Obviously, in the theory under
consideration this behavior is determined by the dependence of the
vertex $v(p,p')$ on momenta. In leading order we can restrict
ourselves to the Lagrangian  (\ref{lagr}) and correspondingly to
the vertex $v(p,p')=C_0$. This means that the $T$ matrix must be of
the form
\begin{equation}
\langle{\bf p}_2|T^{(0)}(z)|{\bf p}_1\rangle=t(z),
\end{equation}
i.e., this is the separable case with the form factor $\psi({\bf
p})=1$. Taking into account that in this case $\alpha=0$, from
Eq. (32), we get
\begin{eqnarray}
\langle{\bf p}_2|T^{(0)}(z)|{\bf
p}_1\rangle=-\frac{b_1^2}{b_2-ib_1\sqrt{z}}\nonumber\\
=-\left[\frac{1}{C_0^{(R)}}+\frac{im^{
\frac{3}{2}}{\sqrt{z}}}{4\pi}\right]^{-1},
\end{eqnarray}
with $b_1=-\frac{4\pi}{m\sqrt{m}}$ and
$C_0^{(R)}=\frac{b_1^2}{b_2}$. Thus the requirement that the T
matrix being of the form (41) satisfies Eq. (\ref{dif}) determines
it up to one arbitrary parameter $C_0^{(R)}$, and we get the
expression (40) which in the EFT is obtained, by summing the
bubbles diagrams, and represents the leading order contribution to
the $T$ matrix. The generalized interaction operator corresponding
to the solution  (42) is of the form
\begin{eqnarray}
 \langle{\bf p}_2| {H}^{(s)}_{int}(\tau)|{\bf p}_1\rangle = \frac{4\pi
 i}{m^\frac{3}{2}\sqrt{\tau}}\exp\left(\frac{i\pi}{4}\right)+
 \frac{16\pi^2}{m^{3}C_0^{(R)}}.\label{zero}
\end{eqnarray}
Thus the renormalization of the EFT in leading order, i.e. of the
theory with the Lagrangian (39), gives rise to the dynamics
governed by the generalized dynamical Eq. (\ref{main}) with
the nonlocal-in-time interaction operator (\ref{zero}). This
operator is a particular case of the interaction operator  (31) of
our model considered in Sec. III. Note also that, as has been shown
in [17], the $T$ matrix obtained in Ref. [19], by using the
dimensional regularization of a model with the separable potential
$V({\bf p}_2,{\bf p}_1)=\lambda\varphi^*({\bf p}_2)\varphi({\bf
p}_1)$, where $\varphi({\bf p})=(d^2+p^2)^{-\frac{1}{4}}$,
satisfies the generalized dynamical equation with the
nonlocal-in-time interaction operator (34).

In leading order we have proved the fact that after
renormalization the dynamics of the EFT under consideration is
governed by the generalized dynamical Eq. (\ref{main}) with a
nonlocal-in-time interaction operator. We hope that this fact can
be proved in any order of the expansion in the number of
derivatives at the vertices or particle lines, and in general one
can expect that renormalization gives rise to the fact that the
dynamics of a quantum system is governed  by the generalized
dynamical Eq. (\ref{main}) with a nonlocal-in-time
interaction operator. In Ref. [16] this fact has been shown, by
using the example of the three-dimensional theory of a neutral
scalar field interacting through a $\varphi^4$ coupling.  The
above gives reason to suppose that such a dynamical situation
takes place after renormalization in any  theory, for example, in
EFT's. Below we will present some general arguments leading to
this conclusion.

Let $G_\Lambda(z)$ be the Green operator of a renormalizable
theory corresponding to the momentum cutoff interaction
Hamiltonian $H_I^{(\Lambda)}(t)$ with renormalized constants. For
every finite cutoff $\Lambda$, the operator $G_\Lambda(z)$
obviously satisfies the Hilbert identity (\ref{green})
\begin{equation}
G_\Lambda(z_1)-G_\Lambda(z_2)=(z_2-z_1)G_\Lambda(z_2)G_\Lambda(z_1).
\label{grl}
\end{equation}
At the same time, the renormalized Green operator $G_{ren}(z)$ is
a limit of the consequence of the operators $G_\Lambda(z)$ for
$\Lambda\to\infty$. Since Eq. (\ref{grl}) is satisfied for every
$\Lambda$, and contains only the operator $G_\Lambda(z)$, the
renormalized Green operator must also satisfy the equation
$$
G_{ren}(z_1)-G_{ren}(z_2)=(z_2-z_1)G_{ren}(z_2)G_{ren}(z_1),
$$
despite the fact that the renormalized Green operator cannot be
represented in the form (\ref{reso}) [in the limit
$\Lambda\to\infty$ the operators $H_I^{(\Lambda)}(t)$ do not
converge to some operator acting on the Hilbert space].
Correspondingly the renormalized $T$ matrix satisfies Eqs.
(\ref{dif}) and (\ref{difer}), despite the fact that in this case
the $LS$ and Schr{\"o}dinger equations do not follow from these
equations. Here the advantages of the GQD manifest themselves.
Within the GQD the dynamical Eq. (\ref{main}) is derived as a
consequence of the most general physical principles, and for Eq.
(\ref{difer}) to be satisfied, the Green operator $G(z)$ need not
be represented in the form (\ref{reso}). In the GQD this operator
is defined by Eq. (\ref{lventa}), where the $T$ matrix in turn is
defined by Eq. (\ref{schr}), i.e., is expressed in terms of the
amplitudes $\langle\psi_2|\tilde S(t_2,t_1)|\psi_1\rangle$ being
the contributions to the evolution operator from the processes in
which the interaction in a quantum system begins at time $t_1$ and
ends at time $t_2$. Only in the case where the interaction
operator is of the form  (\ref{delta}), i.e., the dynamics of the
system is Hamiltonian, can the operator $G(z)$  be represented in
the form  (\ref{reso}). For every finite cutoff $\Lambda$ the
dynamics of the system is Hamiltonian. At the same time, in the
limiting case $\Lambda\tend\infty$ the dynamics is governed by the
generalized dynamical equation with a nonlocal-in-time interaction
operator, i.e., the dynamics is not Hamiltonian.

\section{Effects of the nonlocality of the $NN$ interaction on the
character of nucleon dynamics}

As we have shown, after renormalization the dynamics of the theory
with Lagrangian (39), i.e., the EFT in leading order, is governed
by the generalized dynamical Eq. (\ref{main}) with
a nonlocal-in-time interaction operator (43). This dynamics is the
same as in the model considered in Sec. III with $\alpha=0$. As has
been shown in Ref. \cite{R.Kh.:1999} this is the case when the
dynamics of a quantum system is not Hamiltonian. In order to
clarify this point let us consider the specific features of the
evolution operator in the nonlocal case $\alpha\leq\frac{1}{2}$.
In the Schr{\"o}dinger picture, the evolution operator
$$
\langle{\bf p}_2|V(t)|{\bf p}_1\rangle\equiv\langle{\bf
p}_2|U_s(t,0)|{\bf p}_1\rangle$$ of the theory with the
interaction operator (31) can be rewritten in the form
\begin{eqnarray}
\langle{\bf p}_2|V(t)|{\bf p}_1\rangle =\langle{\bf p}_2|{\bf
p}_1\rangle \exp(-iE_{p_2}t)\nonumber\\+ \frac {i}{2\pi}
\int_{-\infty}^{\infty} dx \frac {\exp(-izt) \langle{\bf
p}_2|T(z)|{\bf p}_1\rangle} {(z-E_{p_2})(z-E_{p_1})},
\end{eqnarray}
where $\langle{\bf p}_2|T(z)|{\bf p}_1\rangle$ is given by Eq.
(\ref{nz}). Since this $T$ matrix satisfies Eqs.  (\ref{dif}) and
(\ref{difer}), the evolution operator (33) is unitary, and
satisfies the composition law (\ref{compos}). Correspondingly, the
operator $V(t)$ constitutes a one-parameter group of unitary
operators, with the group property
\begin{equation}
V(t_1+t_2) = V(t_1) V(t_2) , \quad    V(0)= {\bf 1}.
\end{equation}
Assume that this group has a self-adjoint infinitesimal generator
$H$ which in the Hamiltonian formalism is identified with the
total Hamiltonian. Then for $|\psi\rangle\in{\cal D}(H)$ we have
\begin{equation}
\frac{V(t)|\psi\rangle - |\psi\rangle}{t}\tend\limits_{t \tend
0}-iH|\psi\rangle.
\end{equation}
From this and Eq. (45) it follows that
$$H=H_0+H_I,$$
with
\begin{eqnarray}
\langle{\bf p}_2|H_I|{\bf p}_1\rangle= \frac {i}{2\pi}
\int_{-\infty}^{\infty} dx \frac {z\langle{\bf p}_2|T(z)|{\bf
p}_1\rangle} {(z-E_{p_2})(z-E_{p_1})},
\end{eqnarray}
where $z=x+iy$, and $y>0$. Since $\langle{\bf p}_2|T(z)|{\bf
p}_1\rangle$ is an analytic function of $z$, and, in the case
$\alpha\leq\frac{1}{2}$, tends to zero as $|z|\tend \infty$, from
Eq. (48) it follows that $\langle{\bf p}_2|H_I|{\bf p}_1\rangle=0$
for any ${\bf p}_2$ and ${\bf p}_1$, and hence $H=H_0$. This means
that, if the infinitesimal generator of the group of the operator
$V(t)$ exists, then it coincides with the free Hamiltonian, and
the evolution operator is of the form $V(t)=\exp(-iH_0t)$. Thus,
since this, obviously, is not true, the group of the operators
$V(t)$ has no infinitesimal generator, and hence the dynamics is
not governed by the Schr{\"o}dinger equation.

It should be also noted that in the case $\alpha\leq\frac{1}{2}$,
$\tilde S(t_2,t_1)$ is not an operator on the Hilbert space. In
fact, the wave function
\begin{eqnarray}
\psi({\bf p})\equiv\langle{\bf p}|\psi\rangle=\langle{\bf
p}|\tilde S(t_2,t_1)|\psi_1\rangle
\nonumber \\
=\varphi^*({\bf p})\tilde s(t_2,t_1)\int \frac{d^3k}{(2\pi)^3}
\varphi({\bf k})\langle{\bf k}|\psi_1\rangle
\end{eqnarray}
is not square integrable for any nonzero $|\psi_1\rangle$, because
of the slow rate of decay of the form factor $\varphi({\bf p})$ as
$|{\bf p}|\tend\infty$.  Correspondingly, in the case $\alpha\leq
\frac{1}{2}$ the $T$ matrix given by Eq. (32) does not represent
an operator on the Hilbert space. However, as we have stated, in
general $\tilde S(t_2,t_1)$ may be only an operator-valued
generalized function such that the evolution operator is an
operator on the Hilbert space. Correspondingly, the $T$ matrix
must be such that $G(z)$ given by Eq. (12) is an operator on the
Hilbert space. The $T$ matrix and $\tilde S (t_2,t_1)$ satisfy
these requirements not only for $\alpha>\frac{1}{2}$ but also for
$-\frac{1}{2}<\alpha\leq\frac{1}{2}$, since the evolution operator
given by Eq. (33) and the corresponding $G(z)$ are operators on
the Hilbert space in this case. At the same time, in the case
$\alpha\leq\frac{1}{2}$ we go beyond Hamiltonian dynamics. Thus
the dynamics generated by the EFT is not governed by the
Schr{\"o}dinger equation, and the $NN$ interaction cannot be
parametrized by an interaction Hamiltonian defined on the Hilbert
space. On the other hand, this dynamical situation is not peculiar
from the point of view of the GQD: The Schr{\"o}dinger equation is
only a particular case of the generalized dynamical Eq.
(\ref{main}) where the interaction is instantaneous, and the above
means that the $NN$ interaction is nonlocal in time. It is
extremely important that within GQD the EFT description of low
energy nucleon dynamics becomes as well founded as the
nonrelativistic quantum mechanics describing atomic phenomena. In
fact, in both cases the dynamics is governed by the generalized
dynamical equation (\ref{main}), and only the interaction
operators are different.

Since the interaction operator (43) represents only the contact
component of the $NN$ interaction, in general one has to consider it
in combination with a long-range component of the interaction. In
this case the interaction operator is of the form
\begin{eqnarray}
 \langle{\bf p}_2| {H}^{(s)}_{int}(\tau)|{\bf p}_1\rangle =
 -2i\delta(\tau) V({\bf p}_2,{\bf p}_1)\nonumber\\+
 \langle{\bf p}_2| {H}_{non}(\tau)|{\bf p}_1\rangle, \label{operator}
\end{eqnarray}
where $ \langle{\bf p}_2| {H}_{non}(\tau)|{\bf p}_1\rangle$
represents the contact nonlocal-in-time component, and  $V({\bf
p}_2,{\bf p}_1)$ is a potential describing the long-range
component of the $NN$ interaction.  In leading order the nonlocal
component is given by Eq. (43), and, for the interaction operator,
we can write
\begin{eqnarray}
 \langle{\bf p}_2| {H}^{(s)}_{int}(\tau)|{\bf p}_1\rangle = \frac{4\pi
 i}{m^\frac{3}{2}\sqrt{\tau}}\exp\left(\frac{i\pi}{4}\right) +
 \frac{16\pi^2}{m^{3}C_0^{(R)}}\nonumber\\
 -2i\delta(\tau)V({\bf p}_2,{\bf p}_1). \label{oper}
\end{eqnarray}
The dynamical Eq. (\ref{dif}) with the interaction operator
(51) does not require regularization and renormalization and is as
convenient for numerical calculations as the $LS$ equation. However,
in some cases it is convenient to reduce it to integral equations.
For example, for some potentials  the solution of the dynamical
Eq. (\ref{dif}) can be represented (see the Appendix) in the
form
\begin{eqnarray}
 \langle{\bf p}_2| T(z)|{\bf p}_1\rangle =t_0(z)+t_1(z;{\bf p}_1)+
 t_1(z;{\bf p}_2)\nonumber\\
 +t_2(z;{\bf p}_2,{\bf p}_1), \label{Tmatrix}
\end{eqnarray}
where $t_2(z;{\bf p}_2,{\bf p}_1)$ is a solution of the equation
\begin{eqnarray}
t_2(z;{\bf p}_2,{\bf p}_1)=V({\bf p}_2,{\bf p}_1)\nonumber\\ +\int
\frac{d^3q}{(2\pi)^3}\frac{K(z;{\bf q},{\bf
p}_2)}{z-E_q}t_2(z;{\bf q},{\bf p}_1),\label{t3}
\end{eqnarray}
with
\begin{eqnarray}
K(z;{\bf q},{\bf p}_2)=N(z)\int \frac{d^3k}{(2\pi)^3}\frac{V({\bf
p}_2,{\bf k})}{z-E_k} +V({\bf q},{\bf p}_2),
\end{eqnarray}
and the functions $t_0(z)$ and $t_1(z;{\bf p})$ are defined as
\begin{eqnarray}
t_1(z;{\bf p})=N(z)\int \frac{d^3k}{(2\pi)^3}\frac{t_2(z;{\bf
p},{\bf q})}{z-E_k},\label{t1}
\end{eqnarray}
\begin{eqnarray}
t_0(z)=N(z)\left(1+\int \frac{d^3q}{(2\pi)^3}\frac{t_1(z;{\bf
q})}{z-E_q}\right),\label{t0}
\end{eqnarray}
where
$$N(z)=-\left(\frac{1}{C_0^{(R)}}+\frac{im^{
\frac{3}{2}}{\sqrt{z}}}{4\pi}\right)^{-1}.$$
Equation (53) is a
generalization of the $LS$ equation to the case where the
interaction operator contains a nonlocal-in-time component.

In general the long-range component of the $NN$ interaction consists
of the meson-exchange potentials and the Coulomb potential in the
proton-proton channel. Since the contact term in Eq. (51) is of the
leading order it is natural to keep in $V({\bf p}_2,{\bf p}_1)$
the components of the same order. In Weinberg's power counting the
one-pion-exchange potential is of the leading order. Hence in this
order the $NN$ interaction operator can be expressed as
\begin{eqnarray}
 \langle{\bf p}_2| {H}^{(s)}_{int}(\tau)|{\bf p}_1\rangle = \frac{4\pi
 i}{m^\frac{3}{2}\sqrt{\tau}}\exp\left(\frac{i\pi}{4}\right)+
 \frac{16\pi^2}{m^{3}C_0^{(R)}}\nonumber\\
 -2i\delta(\tau)V_\pi({\bf p}_2,{\bf p}_1), \label{pioper}
\end{eqnarray}
where $V_\pi({\bf p}_2,{\bf p}_1)$ is the conventional
one-pion-exchange potential. Substituting this potential into
Eq. (\ref{dif}) and solving it numerically, one can easily obtain
the $T$ matrix and hence the evolution operator. Note that the
conventional way of solving the above problem is the formal use of
the potential [7,11]
\begin{eqnarray}
V_0({\bf p}_2,{\bf p}_1)=\widetilde{C}+V_\pi({\bf p}_2,{\bf
p}_1).\label{widetilde}
\end{eqnarray}
We say "formal" since the use of such a potential leads to UV
divergencies, and the Schr{\"o}dinger and $LS$ equations require
regularization and renormalization. On the other hand, as we have
shown, the contact interaction, which in Eq. (\ref{widetilde}) is
formally represented by the term $\widetilde{C}$, is parametrized
by the operator (44)  [the first two terms of the operator
(\ref{pioper})]. In this case we deal with  well-defined
interaction operators and Eq. (\ref{dif}), which do not require
regularization and renormalization. By using Eq. (\ref{dif}), one
can obtain the $T$ matrix as easily as in the case of the pure
one-pion-exchange potential.

Thus the formalism of the GQD allows one to formulate an EFT
theory of nuclear forces as a self-consistent theory free from UV
divergences. One of the advantages of such a formulation is that
one can investigate the general consequences of the theory that
have significant effects on the character of nucleon dynamics. As
is well known, the dynamics of many nucleon systems depends on the
off-shell properties of the two-nucleon amplitudes. For this
reason, let us consider the effects of the nonlocality of the $NN$
interaction on these properties. In the nonlocal case, the matrix
elements of the evolution operator as functions of momenta do not
go to zero at infinity as fast as  is required by ordinary quantum
mechanics, and within Hamiltonian formalism this leads to
ultraviolet divergences. For example, in this case the two-nucleon
amplitudes $\langle{\bf p}_2|T(z)|{\bf p}_1\rangle$  do not go to
zero fast enough to make the Faddeev equation well behaved.
 Another consequence of nonlocality in time of the $NN$ interaction is that
for fixed momenta ${\bf p}_1$ and ${\bf p}_2$ the matrix elements
$\langle{\bf p}_2|T(z)|{\bf p}_1\rangle$ tend to zero as $|z|\to
\infty$, while, in the local case, they tend to $\langle{\bf
p}_2|V|{\bf p}_1\rangle$ in this limit. To illustrate this, we
present in Fig. 4 the off-shell behavior of $\langle{\bf
p}_2|T(z)|{\bf p}_1\rangle$ in the limit $|z|\to \infty$. Thus,
the nonlocality in time of the $NN$ interaction  gives rise to an
anomalous off-shell behavior of the two-nucleon amplitudes. The
off-shell properties of the amplitudes for the ordinary
interaction operator and the operator containing the nonlocal term
are qualitatively different. This is true even when the two
interaction operators have approximately the same phase shifts.

\begin{figure}
\resizebox{1\columnwidth}{!} {\includegraphics{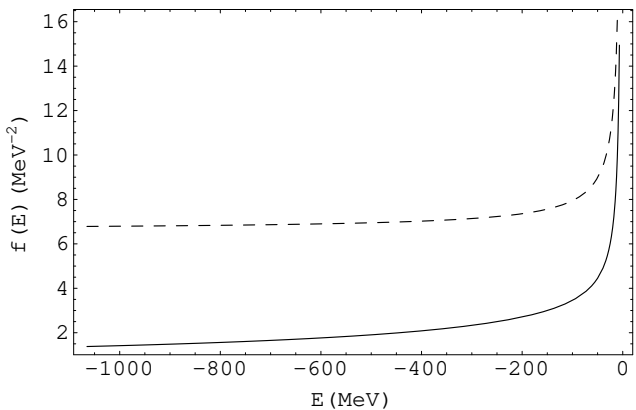}} \caption{The
behavior of the off-shell  amplitude $f(E)=10^9 \langle{\bf
p}|T(z)|{\bf p}\rangle$ ($|{\bf p}|=500$ MeV) in the ${}^3S_1$
channel for $np$ scattering. The solid curve corresponds to the
model with a generalized interaction operator (\ref{less}),
compared to the model with Yamaguchi potential with parameters
given in Ref. \cite{Yam} (dashed line).}
\end{figure}

\section{Summary and discussion}

In leading order of the EFT approach we have shown that the
effective $NN$ interaction is nonlocal in time, and low energy
nucleon dynamics is governed by the dynamical Eq.
(\ref{main}) with the nonlocal-in-time interaction operator (43).
Thus, in the case of quantum mechanics of nucleons at low
energies, we deal with the non-Hamiltonian dynamics that has been
investigated in Ref. \cite{R.Kh.:1999}. This dynamics is described
in a natural way by the GQD that, as has been shown, provides a
consistent parametrization of the $NN$ interaction.  The leading
order contact component of this interaction is parametrized by the
nonlocal-in-time operator (43). This operator is well defined, and
the generalized dynamical equation with this operator does not
requires regularization and renormalization. The parametrization
of the  $NN$ interaction that takes into account the long-range
component is represented by Eq. (51).

The fact that in leading order of the EFT approach low energy
nucleon dynamics is described by the model considered in Sec. III
allows us to explain the advantages of the GQD in describing
nucleon dynamics in terms of this model. As we have shown, the
generalized dynamical equation permits any form factor
$\varphi({\bf p})$ in Eq. (20) having UV behavior (19) with
$\alpha>-\frac{1}{2}$. In the case $\alpha>\frac{1}{2}$, the only
possible form of the interaction operator is Eq. (26). In this
case we deal with the ordinary separable-potential model. In the
case $-\frac{1}{2}<\alpha<\frac{1}{2}$, the interaction operator
must be of the form (31), and hence the interaction generating the
dynamics of the system is necessarily nonlocal in time. In this
case the $T$ matrix is given by Eq. (32). On the other hand, one
can obtain the same $T$ matrix in another way, starting with the
singular potential
\begin{eqnarray}
 V({\bf p}',{\bf p})=\lambda \varphi^*({\bf p}')
\varphi({\bf p}),\quad \varphi({\bf p})\sim|{\bf
p}|^{-\alpha},\quad\alpha\leq\frac{1}{2} \label{pot}
\end{eqnarray}
that does not make sense without renormalization, since it leads
to UV divergencies. By solving the $LS$ equation regulated in some
way and performing renormalization, one can get expression (32)
for the $T$ matrix. However, in this way one cannot determine a
potential describing the fundamental interaction in the system.
The singular potential (\ref{pot}) is only of formal importance
for the problem under consideration. All the information contained
in this potential is that the $T$ matrix is of the form
$\langle{\bf p}_2|T(z)|{\bf p}_1\rangle=\varphi^*({\bf
p}_2)\varphi({\bf p}_1)t(z)$ with the same form factor. If the
group of the evolution operators given by Eq. (45) had an
infinitesimal generator, one could identify it with the
Hamiltonian. However, as we have shown in Sec. V, this group has
no infinitesimal generator, and hence there are no potentials that
could  govern the dynamics of the system after renormalization.
Moreover, the $T$ matrix (32) does not satisfy the $LS$ equation
and has such properties that are at variance with the Hamiltonian
formalism. Thus in this case we have only a calculation rule that
allows one to compute results for physical quantities. On the
other hand, the above problems are the cost of trying to describe
the dynamics of the system after renormalization in terms of the
Hamiltonian formalism, despite the fact that this dynamics is
non-Hamiltonian. From the more general point  of view provided by
the GQD we see that in this case the $T$ matrix satisfies the
generalized dynamical Eq. (\ref{dif}) with the nonlocal-in-time
interaction operator (31), and this operator describes the
fundamental interaction generating the dynamics of the system.
Thus after renormalization in the theory with the potential
(\ref{pot}) we have the dynamics that, according to the GQD, must
take a place in the case $-\frac{1}{2}<\alpha<\frac{1}{2}$, and an
example of such a theory is low energy nucleon dynamics in leading
order where the singular potential is $V({\bf p}_2, {\bf
p}_1)=\widetilde{C}$. Within the GQD this theory is internally
consistent and is as well founded as theories with ordinary
potentials such as quantum mechanics describing atomic phenomena.
For example, the $T$ matrix is well defined, and its properties
satisfy the general requirements of the theory. This is  not only
important from the point of view of the internal consistency of
the theory. Only a well-founded theory provides the possibility to
obtain, in a theoretical way, new knowledge, and to prove the
correctness of calculations performed within its framework. The
fact that in the approach based on the GQD we have the
well-defined dynamical equation is a great advantage of this
approach for practical calculations.  This equation does not
require regularization and renormalization and is as convenient
for numerical calculations as the $LS$ equation.

It is important that the use of the GQD for investigating the
dynamical situation in the EFT of nuclear forces gives rise to a
natural parametrization of the $NN$ interaction, and uniquely
determines the form of the generalized interaction operator
describing the interaction of nucleons. In contrast with initial
Lagrangians of EFT's or formal singular potentials that are
produced by these Lagrangians, knowing the generalized operator of
the $NN$ interaction is sufficient to compute results for physical
quantities. We have obtained that in leading order the contact
term of the $NN$ interaction is parametrized by the interaction
operator (43). In the same order (in Weinberg's power counting)
the $NN$ interaction is parametrized by the operator (\ref{pioper}).
In this way one can construct the interaction operators in any
order of the EFT approach. This operator can then be used, for
example, for determining the interaction operators parametrizing
interactions of nuclei.

\appendix
\section{}

 Let us consider the solution of Eq. (\ref{dif}) in the case where the
 interaction operator is of the form (51). From Eqs. (14) and (15) it
 follows that this solution can be
 represented in the form
 \begin{equation}
\langle{\bf {p}}_2|T(z)|{\bf {p}}_1 \rangle= \lim \limits_{u \tend
-\infty}\langle{\bf {p}}_2|T_u(z)|{\bf {p}}_1 \rangle,
\end{equation}
 where the operator $T_u(z)$ is the solution of the equation
 \begin{equation}
T_u(z)=B(u)+(u-z)B(u)G_0(u)G_0(z)T_u(z).\label{equ}
\end{equation}
Here the operator $B(z)$ is given by
\begin{eqnarray}
 \langle{\bf p}_2| B(z)|{\bf p}_1\rangle =f_1(z)
 +V({\bf p}_2,{\bf p}_1),
\end{eqnarray}
with $$f_1(z)=-\frac{4\pi}{m^\frac{3}{2}\sqrt{-z}}-
 \frac{16\pi^2}{m^{3}C_0^{(R)}z}.$$
 The solution of Eq. (A2) can be represented in the form
\begin{eqnarray}
\langle{\bf {p}}_2|T_u(z)|{\bf {p}}_1
\rangle=t_0^{(u)}(z)+t_1^{(u)}(z;{\bf {p}}_1)\nonumber\\+
\widetilde{t}_1^{(u)}(z;{\bf {p}}_2)+t_2^{(u)}(z;{\bf {p}}_1,{\bf
{p}}_2).
\end{eqnarray}
 Substituting this representation in Eq. (\ref{equ})
yields the following equations for $t_0^{(u)}(z)$,
$t_1^{(u)}(z;{\bf {p}})$, $\widetilde{t}_1^{(u)}(z;{\bf {p}}_2)$
and $t_2^{(u)}(z;{\bf {p}}_1,{\bf {p}}_2)$:
\begin{eqnarray}
t_0^{(u)}(z)=f_1(u)+(u-z)f_1(u)\nonumber\\
\times\int
\frac{d^3k}{(2\pi)^3}\frac{\left(t_0^{(u)}(z)+t_1^{(u)}(z;{\bf
k})\right)} {(z-E_k)(u-E_k)},\label{tu0}
\end{eqnarray}
\begin{eqnarray}
 t_1^{(u)}(z;{\bf p})=(u-z)\int
\frac{d^3k}{(2\pi)^3} \frac{\left(t_0^{(u)}(z)+t_1^{(u)}(z;{\bf
k})\right)}
{(z-E_k)(u-E_k)}\nonumber\\
\times V({\bf k},{\bf p}), \label{tu1}
\end{eqnarray}
\begin{eqnarray}
 \widetilde{t}_1^{(u)}(z;{\bf p})=(u-z)f_1(u) \nonumber\\
\times\int
\frac{d^3k}{(2\pi)^3}\frac{\left(\widetilde{t}_1^{(u)}(z;{\bf
p})+t_2^{(u)}(z;{\bf k},{\bf p})\right)} {(z-E_k)(u-E_k)},
\end{eqnarray}
\begin{eqnarray}
t_2^{(u)}(z;{\bf p}_2,{\bf p}_1)=V({\bf p}_2,{\bf
p}_1)+(u-z)\int \frac{d^3k}{(2\pi)^3}\nonumber\\
\frac{\left(\widetilde{t}_1^{(u)}(z;{\bf p}_2)+t_2^{(u)}(z;{\bf
p}_2,{\bf k})\right)} {(z-E_k)(u-E_k)}V({\bf k},{\bf p}_1).
\label{tu3}
\end{eqnarray}
 It is not difficult to verify that
 $$t_1^{(u)}(z;{\bf p})= \widetilde{t}_1^{(u)}(z;{\bf p}).$$
By solving the above set of equations, one can obtain  the
functions $t_0^{(u)}(z)$, $t_1^{(u)}(z,{\bf{p}})$ and
$t_2^{(u)}(z,{\bf{p}}_2,{\bf{p}}_1)$ that in turn can be used for
constructing the $T$ matrix. In fact, from Eqs. (A1) and (A2) it follows
that  the $T$ matrix can be represented in the form (\ref{Tmatrix})
where the functions $t_0(z)$, $t_1(z,{\bf{p}})$ and
$t_2(z,{\bf{p}}_2,{\bf{p}}_1)$ are given by
\begin{eqnarray}
t_0(z)=\lim_{u\to-\infty}t_0^{(u)}(z),\quad t_1(z,{\bf
{p}})=\lim_{u\to-\infty}t_1^{(u)}(z,{\bf {p}}),\nonumber
\\
t_2(z,{\bf {p}}_2,{\bf {p}}_1)=\lim_{u\to-\infty} t_2^{(u)}(z,{\bf
{p}}_2,{\bf {p}}_1).\nonumber
\end{eqnarray}
Taking into account that
$$
\int
\frac{d^3k}{(2\pi)^3}\frac{u-z}{(z-E_k)(u-E_k)}=\frac{m\sqrt{m}}{4\pi}
\left(\sqrt{-z}-\sqrt{-u}\right),
$$
equation (\ref{tu0}) can be rewritten in the form
\begin{eqnarray}
t_0^{(u)}(z)=-\frac{4\pi}{m^\frac{3}{2}
\sqrt{-u}}+t_0^{(u)}(z)\nonumber\\
\times\left(-\frac{4\pi}{m^\frac{3}{2}\sqrt{-u}}-\frac{16\pi^2}{m^3C_0^{(R)}u}\right)
\frac{m\sqrt{m}}{4\pi}
\left(\sqrt{-z}-\sqrt{-u}\right)\nonumber\\
-\frac{4\pi}{m^\frac{3}{2}\sqrt{-u}}\int d^3 k\frac{t_1(z,{\bf
{k}})}{z-E_k}+o(|u|^{-1/2}).\nonumber
\end{eqnarray}
Letting $u\to -\infty$ in this equation and assuming that $V({\bf
p}_2,{\bf p}_1)$ goes to zero sufficiently fast when momenta tend
to infinity, one can easily get Eq. (\ref{t0}). In the same way,
from Eqs. (\ref{tu1}) and (\ref{tu3}) one can derive Eqs.
(\ref{t3}) and (\ref{t1}).

\end{document}